\documentclass{article}
\usepackage{spconf,amsmath,graphicx}
\usepackage{preamble}

\usepackage[skip=8pt]{caption}
\usepackage{comment}

\title{Light-SERNet: A lightweight fully convolutional neural network for speech emotion recognition}
\name{Arya~Aftab$^{1,2}$, Alireza~Morsali$^{3}$, Shahrokh Ghaemmaghami$^{1,2}$, Benoit Champagne$^{3}$ 
}
\address{$^{1}$ 
Department of Electrical Engineering, Sharif University of Technology, Tehran, Iran
\\$^{2}$
Electronics Research Institute, Sharif University of Technology, Tehran, Iran
\\$^{3}$ 
Department of Electrical and Computer Engineering, McGill University, Montreal, Canada\\
\normalsize aftab.arya@ee.sharif.edu,  alireza.morsali@mail.mcgill.ca, ghaemmag@sharif.edu,   benoit.champagne@mcgill.ca}
\begin{document}
%
\maketitle
\begin{abstract}
  Detecting emotions directly from a speech signal plays an important role in effective human-computer interactions. Existing speech emotion recognition models require massive computational and storage resources, making them hard to implement concurrently with other machine-interactive tasks in embedded systems.
  In this paper, we propose an efficient and lightweight fully convolutional neural network for speech emotion recognition in systems with limited hardware resources. In the proposed FCNN model, various feature maps are extracted via three parallel paths with different filter sizes. This helps deep convolution blocks to extract high-level features, while ensuring sufficient separability. The extracted features are used to classify the emotion of the input speech segment. While our model has a smaller size than that of the state-of-the-art models, it achieves a higher performance on the IEMOCAP and EMO-DB datasets. The source code is available {\color{purple}\url{https://github.com/AryaAftab/LIGHT-SERNET}}
\end{abstract}
\begin{keywords}
Speech emotion recognition, lightweight model, convolutional neural network, Mel frequency Cepstrum coefficient (MFCC)
\end{keywords}
\section{Introduction}
\label{sec:intro}
Detecting emotions directly from a speech signal plays an important role in effective human-computer interactions \cite{pandey2019deep}. Automatic emotion recognition can be potentially used in a wide range of smart devices, especially in intelligent dialogue systems and voice assistants, such as Apple Siri, Amazon Alexa, and Google Assistant. Recently, identifying the emotional state of speakers from their speech utterances have received considerable attention \cite{han2014speech, li2019dilated, zhong2020lightweight, chen20183, yenigalla2018speech, satt2017efficient, zhao2019compact}.
Existing benchmarks of speech emotion recognition (SER) methods are mainly comprised of a feature extractor and a classifier to obtain the emotional states \cite{han2014speech}.


Recently, deep learning (DL) based techniques have revolutionized the field of speech processing and in many cases outperformed classical methods \cite{han2014speech, deep_benefit1}. One of the main reasons for the success of DL-based methods is the ability of deep neural networks (DNNs) to extract complex features from the data through a learning process \cite{lecun2015deep}.


In particular, convolutional neural networks (CNNs) have achieved significant improvements in SER, as compared to conventional methods \cite{cnn3, cnn1, cnn2}.
CNNs are particularly powerful for disregarding the information conveyed by the input signal that could be irrelevant to the target task  \cite{cnn_effect}. This characteristics is especially useful when the input is a complex unstructured signal, such as an image or a speech signal.
Yenigalla et al. \cite{yenigalla2018speech} increased the recognition rate by using several parallel paths with large convolutional filters and phoneme embedding.
Chen et al. \cite{chen20183} used Mel-spectrogram, deltas, and delta-deltas as inputs and proposed a 3-D attention-based convolutional recurrent neural network to preserve effective emotional information and reducing the influence of irrelevant emotional factors.
Li et al. \cite{li2019dilated} proposed a combination of dilated residual network and multi-head self-attention to relieve the loss of temporal structure of speech in the progressive resolution reduction, while ignoring relative dependencies between elements in suprasegmental feature sequences.
To reduce the model size and computational costs, Zhong et al. \cite{zhao2019compact} quantized the weights of the neural networks from the original full-precision values into binary values that can then be stored and processed more easily.
Zhong et al. \cite{zhong2020lightweight} combined the attention mechanism and the focal loss, which concentrate the training process on learning hard samples and down-weighing easy samples, to resolve the problem with challenging samples.

In this paper, we propose a novel model for SER that can learn spectro-temporal information from Mel frequency cepstral coefficients (MFCC), which only make use of fully CNN. First, a hierarchical DL model is developed to automate and replace the process of hand-engineering features. In fact, we take advantage of three parallel CNN blocks to extract features with different properties from MFCC energy maps. The extracted features are then concatenated and fed to a deep CNN to capture high-level representations which are finally classified with a softmax layer. The proposed model is noticeably lightweight which makes it suitable for online SER applications and for implementation on small embedded systems and IoT devices with limited resources. The use of CNNs not only reduces model complexity, but provides better generalization, as compared to that in benchmark methods. Our experiments for evaluation of the proposed SER model, on the IEMOCAP and EMO-DB datasets, 
corroborate that our model requires considerably less parameters, while achieving the same or better performance than that of state-of-the-art models.


\section{Architecture Design} \label{Material}
\begin{figure*}[htp]
\centering
\vspace{-1em}
\includegraphics[scale = 0.81]{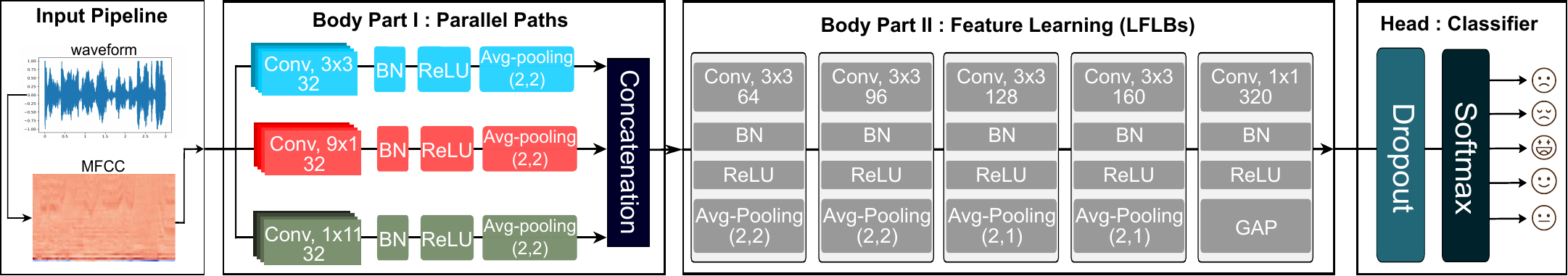}
    \caption{\small The framework of the proposed lightweight fully convolutional neural network for speech emotion recognition.}
    \label{fig:Architecture}
    \vspace{-1em}
\end{figure*}

In this section, we present the proposed architecture which consists of three main parts: input pipeline, feature extraction blocks (Body), and classification block (Head). The body, in turn, consists of two sub-parts: parallel 2D-convolutions (Body Part I) and local feature learning blocks (LFLBs) (Body Part II).
Fig.\ref{fig:Architecture} illustrates structure of the network, whose parts are described in further details below. 



\subsection{Input pipeline} \label{Input_Pipeline}
After normalizing the audio signals between $-1$ and $1$, the MFCCs of the signals are calculated.  To this end, we use a Hamming window to split the audio signal into $64$-ms frames with $16$ms overlaps, which can be considered as quasi-stationary segments. Following a 1024-point Fast Fourier transform (FFT) applied to each frame, the signal undergoes a Mel scale filter bank analysis, in the range of 40 Hz to 7600 Hz. The MFCCs of each frame are then calculated using an inverse discrete cosine transform, where the first 40 coefficients are selected to train the model.

\subsection{Body Part I}
In Body Part I, three parallel CNNs are applied to the MFCC to extract time and frequency features. This structure can achieve a balance between spectral and temporal information in its feature extractor.

In \cite{araujo2019computing}, a direct relationship has been observed between the classification accuracy and receptive field size, which means having a larger receptive field can improve the classification accuracy. Consequently, we use the following techniques to increase the receptive field of a convolution network: 1) increasing the number of layers (deeper network), 2) using sub-sampling blocks such as pooling or higher stride, 3) employing dilated convolutions, and 4) performing depth-wise convolutions.

Deeper networks have higher receptive fields, because each additional layer increases the receptive field by the kernel size \cite{luo2016understanding}. However, increasing the number of layers increases the number of model parameters, which leads to over-fitting of the model.

\begin{figure}[t]
				\centering

				\includegraphics[scale = 0.65]{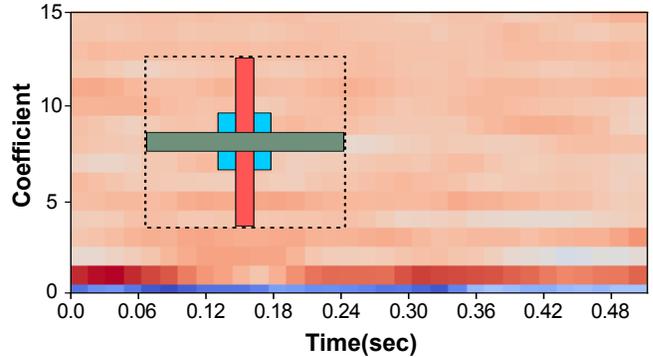}
					\caption{\small The effect of parallel paths (gray, red, and blue rectangles) and their resulting receptive fields (dotted line rectangle).}
					\label{fig:ParralellEffect}
				\end{figure}

For multi-dimensional signals, each dimension can be considered separately for calculating the receptive fields \cite{araujo2019computing}. Hence, we use kernels of size $9\times1$, $1\times11$, and $3\times3$ to extract spectral, temporal, and spectral -temporal dependencies, respectively, as shown in Fig.\ref{fig:ParralellEffect}. The advantage of using this technique over having only one path with the same receptive field size is to reduce the number of parameters and the computational cost of this part of the model by $\frac{9 \times 11}{(9 \times 1 + 1 \times 11 + 3 \times 3)}$. Finally, the extracted features of each path are concatenated and fed into Body II. The second box in Fig.\ref{fig:Architecture} illustrates the structure of Body Part I.
\subsection{Body Part II}
The Body Part II consists of several LFLBs with different configurations applied to the concatenated low-level features from Body part I to capture high-level features.

An LFLB is a collection of successive layers inspired by the work of Zhao et al. \cite{zhao2019speech}. The original LFLB consists of a convolution layer, a batch normalization layer (BN), an exponential linear unit (ELU), and a max-pooling layer. In our work, the ELU layer and the max-pooling layer have been replaced by a rectified linear unit (ReLU) and the average-pooling, respectively.

The last LFLB uses the global average pooling (GAP), instead of the average-pooling, making our model capable of training on datasets of different lengths without changing the architecture. The specifications of the Body Part II are illustrated in Fig.\ref{fig:Architecture}.

\subsection{Head}
The body part is supposed to map the nonlinear input space into a linearly separable sub-space, and thus, one fully-connected layer is enough for the classification. Therefore, the head part includes only a dropout layer to reduce overfitting and a fully-connected layer with a softmax activation function that reduces the computational complexity and the number of parameters, as compared to Attention and LSTM layers.

\section{Experiments and Results} \label{Results}
In this section, we first introduce the datasets, then explain the experimental setup employed to train and evaluate the models, and finally discuss the results and compare them to those of the latest works.

\subsection{Dataset}
To evaluate the proposed model, we use two datasets, namely: the interactive emotional dyadic motion captures (IEMOCAP) \cite{IEMOCAP} and the berlin emotion dataset (EMO-DB) \cite{EMO-DB}. The details of each dataset are given below.\\
\textbf{IEMOCAP}: This multimodal dataset, recorded at the University of Southern California, includes $12$ hours of audio-visual data divided into five sessions, recorded by male and female professional actors and actresses with scripted and improvised scenarios. The scripted part is performed for predetermined emotions, while the improvised part is closer to natural speech. The samples are annotated in both dimensional and categorical forms, which we use only the categorical form. In order to compare the results of the proposed method to those of the previous studies, we first combine the two classes of exciting and happy, and then evaluate the IEMOCAP(scripted+improvised) dataset. The IEMOCAP (improvised) dataset includes $2837$ samples with a class distribution of happiness ($12.3\%$), sadness ($26.9\%$), angry ($12\%$), and natural ($48.8\%$), and the IEMOCAP (scripted+improvised) dataset includes $5531$ samples with a class distribution of happiness ($29.5\%$), sadness ($19.6\%$), angry ($19.9\%$), and natural ($30.8\%$).\\
\textbf{EMO-DB}: This dataset is in German-language, recorded by ten professional actors and actresses (five men and five women). The dataset includes $535$ emotional utterances in $7$ classes: anger ($23.7\%$), natural ($14.7\%$), sadness ($11.5\%$), fear ($12.9\%$), disgust ($8.6\%$), happiness ($13.2\%$) and boredom ($15.1\%$).
\begin{table*}[b]
\caption{\small The proposed model performance of different input lengths between CE-Loss and F-Loss on the IEMOCAP (improvised), IEMOCAP (scripted+improvised), and EMO-DB datasets in terms of UA(\%), WA(\%), and F1(\%).}
\label{tab:overall}
\setlength\tabcolsep{1.5pt}
\setlength\extrarowheight{5pt}
\begin{tabular}{c|l|ccc|ccc|ccc|ccc|ccc|ccc}
\toprule
\multicolumn{2}{c|}{\textbf{}}                                                       & \multicolumn{6}{c|}{\textbf{IEMOCAP(improvised)}}                                                                                                                     & \multicolumn{6}{c|}{\textbf{IEMOCAP(scripted+improvised)}}                                                                                                            & \multicolumn{6}{c}{\textbf{EMO-DB}}                                                                                                                                  \\ \cline{3-20} 
\multicolumn{2}{c|}{\textbf{\begin{tabular}[c]{@{}c@{}}Input\\ Length\end{tabular}}} & \multicolumn{3}{c|}{\textbf{F-Loss}}                                          & \multicolumn{3}{c|}{\textbf{CE Loss}}                                             & \multicolumn{3}{c|}{\textbf{F-Loss}}                                          & \multicolumn{3}{c|}{\textbf{CE Loss}}                                             & \multicolumn{3}{c|}{\textbf{F-Loss}}                                          & \multicolumn{3}{c}{\textbf{CE Loss}}                                             \\ \cline{3-20} 
\multicolumn{2}{c|}{\textbf{}}                                                       & \multicolumn{1}{c|}{\textbf{UA}} & \multicolumn{1}{c|}{\textbf{WA}} & \textbf{F1} & \multicolumn{1}{c|}{\textbf{UA}} & \multicolumn{1}{c|}{\textbf{WA}} & \textbf{F1} & \multicolumn{1}{c|}{\textbf{UA}} & \multicolumn{1}{c|}{\textbf{WA}} & \textbf{F1} & \multicolumn{1}{c|}{\textbf{UA}} & \multicolumn{1}{c|}{\textbf{WA}} & \textbf{F1} & \multicolumn{1}{c|}{\textbf{UA}} & \multicolumn{1}{c|}{\textbf{WA}} & \textbf{F1} & \multicolumn{1}{c|}{\textbf{UA}} & \multicolumn{1}{c|}{\textbf{WA}} & \textbf{F1} \\ \hline
\multicolumn{2}{c|}{\textbf{$3$ seconds}}                                              & $68.37$                            & $77.41$                            & $76.01$       & $68.42$                            & $76.60$                            & $75.44$       & $66.10$                            & $65.47$                            & $65.42$       & $65.81$                            & $65.37$                            & $65.40$       & $92.88$                            & $93.08$                            & $93.05$       & $94.15$                            & $94.21$                            & $94.16$       \\
\multicolumn{2}{c|}{\textbf{$7$ seconds}}                                              & $70.78$                            & $79.87$                            & $78.84$       & $71.51$                            & $78.73$                            & $77.86$       & $70.76$                            & $70.23$                            & $70.20$       & $70.12$                            & $69.15$                            & $69.09$       & -                            & -                            & -       & -                            & -                            & -       \\ 
\bottomrule
\end{tabular}
\vspace{-1em}
\end{table*}
\subsection{Experimental setup}
\textbf{Implementation and training:}
We use the Tensorflow Python Library, version $2.5$, to implement our models. The models are trained on an Nvidia Tesla V$100$ graphical processing unit (GPU) for $300$ epochs and $32$ batch sizes. Adam optimizer with an initial learning rate of $10^{-4}$ is used. The learning rate from epoch $50$ and above decreases by a rate $e^{-0.15}$ every 20 epochs.

\textbf{Regularizers:}
Due to the lack of enough data for training the model, overfitting may be encountered, so we introduce regularization to cope with this problem. We use batch normalization after each convolutional layer, dropout at a rate of $0.3$ before the softmax layer, and weight decay ($L2$ regularization) at a rate of $10^{-6}$ for LFLBs.

\textbf{Metrics:}
As there is data imbalance among classes of datasets, three metrics are used to evaluate the proposed models: 1) unweighted accuracy (UA), 2) weighted accuracy (WA), and 3) F1-score (F1). All the reported experimental results are based on $10$-fold cross-validation.

\textbf{Precision of weights:}
The model weights have $32$-bit floating-point precision during training. Following training the models, we change the precision of the trained model weights to $16$-bit floating-point to reduce the size of the model by half. All reported results are for the weights with this precision.
\begin{table}[t]
    \begin{center}
        \caption{\small Comparison of the model size (MB) and performance with those of other methods, on the IEMOCAP (scripted + improvised), in terms of UA, WA, and F1.}
        \label{tab:table1}
        \setlength\tabcolsep{4.5pt}
        \begin{tabular}{llllll}
        \toprule
        \textbf{Methods} & \textbf{Size} & \textbf{UA(\%)} & \textbf{WA(\%)} & \textbf{F1(\%)} \\ \hline
        Han (2014)\cite{han2014speech} & $12.3$ & $48.20$  & $54.30$ & -  \\
        Li (2019)\cite{li2019dilated} & $9.90$ & $67.40$  & -  & $67.10$  \\
        Zhong (2020)\cite{zhong2020lightweight} & $0.90$ & $71.72$  & $70.39$  & $70.85$  \\
        Ours (F-Loss, $7$sec)  & $0.88$ & $70.76$ & $70.23$ & $70.20$ \\
        \hline
        \end{tabular}
    \end{center}
    \vspace{-1em}
\end{table}

\subsection{Results and Discussions}
\textbf{Impact of loss function:}
We choose two loss functions to train the proposed models: Focal loss (F-Loss) and cross-entropy loss (CE-Loss). F-Loss is presented to address the class imbalance and challenging samples \cite{lin2017focal}. In the experiments, F-Loss with $\gamma=2$ is used. Table \ref{tab:overall} shows the results for the both loss functions on the EMO-DB and IEMOCAP datasets. Comparing to UA in Table \ref{tab:overall}, it is shown that F-Loss achieves higher accuracy than CE-Loss on the IEMOCAP (improvised + scripted), whereas, for the IEMOCAP (improvised) and EMO-DB datasets, CE-Loss performs better. These results indicate that the UA of the models can improve the performance, in some cases, with simple CE-loss (Table \ref{tab:overall}).

\textbf{Impact of parallel paths:}
Here, we evaluate the effect of parallel paths on the IEMOCAP and EMO-DB datasets with CE-Loss. Simultaneous use of the paths has increased the WA, UA, and F1 by $1.38\%$, $0.91\%$, and $1.06\%$, on the IEMOCAP (scripted+improvised) dataset, respectively, as compared to the separate use of the paths. This improvement, on the EMO-DB dataset, has been $1.86\%$, $1.35\%$, and $1.57\%$, respectively. For a fair comparison, the same number of filters have been employed in both the simultaneous and separate uses of paths.

\textbf{Impact of input length:}
Due to the variable length of the IEMOCAP dataset utterances (i.e., in the range of $0.58$ to $34.13$ seconds), we have evaluated the proposed model for input lengths of $3$ and $7$ seconds. The main problem with higher input lengths is the computational cost and peak memory usage (PMU). The computational cost for the input length of $3$ and $7$ seconds is $322$ and $760$ million floating-point operations (MFLOPs), respectively, and the PMU for the input length of $3$ and $7$ seconds is $1610$ and $3797$ kilobytes, respectively. It is also found that using the $7$-second input length instead of the $3$-second input length increases the evaluation metrics on the IEMOCAP (improvised) by more than 2.13\% and the evaluation metrics on the IEMOCAP (scripted+improvised) by more than $3.69\%$ (Table \ref{tab:overall}).

\begin{table}[t]
    \begin{center}
       \caption{\small Comparison of the model size (MB) and performance with those of other methods, on the IEMOCAP (improvised), in terms of UA, WA, and F1.}
        \label{tab:table2}
        \setlength\tabcolsep{4.5pt}
        \begin{tabular}{llllll}
        \toprule
        \textbf{Methods} & \textbf{Size} & \textbf{UA(\%)} & \textbf{WA(\%)} & \textbf{F1(\%)} \\ \hline
        Chen (2018)\cite{chen20183} & $323$ & $64.74$  & -  & -  \\
        Yenigalla(2018)\cite{yenigalla2018speech} & $7.20$ & $61.60$  & $71.30$  & -  \\
        Satt (2017)\cite{satt2017efficient} & $5.50$ & $62.00$  & $67.30$  & -  \\
        Zhao (2019)\cite{zhao2019compact} & $4.34$ & $61.90$  & -  & -  \\
        Ours (F-Loss, $7$sec)  & $0.88$ & $70.78$ & $79.87$ & $78.84$ \\
        \hline
        \end{tabular}
    \end{center}
\end{table}

\textbf{Comparison with state-of-the-art methods:}
Here, we present simulation results to compare our model to several benchmarks on the IEMOCAP (scripted+improvised), IEMOCAP (improvised), and EMO-DB datasets in Tables \ref{tab:table1}, \ref{tab:table2}, and \ref{tab:table3}, respectively.
As shown in Table \ref{tab:table1}, our model has slightly less WA, UA, and F1 than those of the Zhong model~\cite{zhong2020lightweight}, on the IEMOCAP (scripted+improvised) dataset, which can be attributed to model training using different annotations in addition to the label of each utterance. 
On the EMO-DB dataset, due to the unavailability of different annotations for training, our model outperforms the Zhong model~\cite{zhong2020lightweight} by more than $2.4\%$ (Table \ref{tab:table3}). 
As presented in Table \ref{tab:table2}, the proposed model has significant performance improvement on the IEMOCAP (improvised) dataset, as compared to that of the state-of-the-art models. This improvement is achieved while our model is smaller in size, as compared to the state-of-the-art models.

\begin{table}[t]
    \begin{center}
        \caption{\small Comparison of model size (MB) and performance in terms of UA, WA, and F1 with those of other methods on the EMO-DB.}
        \label{tab:table3}
        \setlength\tabcolsep{4.5pt}
        \begin{tabular}{llllll}
        \toprule
        \textbf{Methods} & \textbf{Size} & \textbf{UA(\%)} & \textbf{WA(\%)} & \textbf{F1(\%)} \\ \hline
        Chen (2018)\cite{chen20183} & $323$ & $82.82$ & - & - \\
        Zhao (2019)\cite{zhao2019compact} & $4.34$ & $79.70$  & - & -  \\
        Zhong (2020)\cite{zhong2020lightweight} & $0.90$ & $90.10$ & $91.81$ & $90.67$ \\
        Ours (CE-Loss, $3$sec) & $0.88$ & $94.15$ & $94.21$ & $94.16$ \\
        \hline
        \end{tabular}
    \end{center}
\end{table}

\section{Conclusion}
In this paper we presented an efficient convolutional DNN for speech emotion recongition. 
The proposed fully CNN model extracts various feature maps by means of three parallel paths with different filter sizes. This helps deep convolution blocks to extract high-level features,  while  ensuring  sufficient  separability. These features are finally used for classify the emotions of the speech signal segment.
Comparing to the state-of-the-art models, the proposed model has smaller size to reach almost the same or higher recognition performance. 
\newpage
\bibliographystyle{IEEEtran}
\bibliography{IEEEabrv,Speech_ms}

\end{document}